\documentclass[pre,twocolumn,superscriptaddress,showpacs]{revtex4}

\usepackage{bm, amsmath,amssymb,graphicx}

\def\etal{{\it\,et. al.\,}}
\def\PRL{{\it\,Phys. Rev. Lett.\,}}
\def\PR{{\it\,Phys. Rev.\,}}
\def\JMP{{\it\,Journ. Math. Phys.\,}}
\def\JPA{{\it\,Journ. Phys. {\rm A}\,}}

\begin{document}

\title{Curvature induced quantum potential on deformed surfaces}

\author{Victor Atanasov}
 \altaffiliation[Also at]{ Laboratoire de Physique Th\'{e}orique et
Mod\'{e}lisation , Universit\'{e} de Cergy-Pontoise,
 F-95302 Cergy-Pontoise, France }
\affiliation{Institute for Nuclear Research and Nuclear Energy,
Bulgarian Academy of Sciences,  72 Tsarigradsko chaussee, 1784
Sofia, Bulgaria}
 \email{victor@inrne.bas.bg}

\author{Rossen Dandoloff}
\affiliation{ Laboratoire de Physique Th\'{e}orique et
Mod\'{e}lisation , Universit\'{e} de Cergy-Pontoise,
 F-95302 Cergy-Pontoise, France}
 \email{rossen.dandoloff@ptm.u-cergy.fr}

\begin{abstract}
We investigate the effect of curvature on the behaviour of a
quantum particle bound to move on a surface. For the Gaussian bump
we derive and discuss the quantum potential which results in the
appearance of a bound state for particles with vanishing angular
momentum. The Gaussian bump provides a characteristic length for
the problem. For completeness we propose an inverse problem in differential geometry, i.e. what deformed
surfaces produce prescribed curvature induced quantum potentials.
We solve this inverse problem in the case of rotational surfaces.
We also show that there exist rotational surfaces in the form of a circular
strip around the axis of symmetry which allow particles with
generic angular momentum to bind.
\end{abstract}


\maketitle

The quantum physics of curved surfaces plays an increasingly
significant role in the engineering of devices based on coated
interfaces\cite{1}. Curvature also affects the mechanical
properties of some biological systems\cite{4}. Investigation on
the geometric interaction between defects and curvature in thin
layers of superfluids, superconductors, and liquid crystals
deposited on curved surfaces is under way \cite{Vit*04}. The
curvature of a semiconductor surface determines also an
interesting mechanism of spin--orbit interaction of
electrons\cite{Ent*01}. Goldstone and Jaffe\cite{11} and
Exner\cite{12} proved that at least one bound state exists for all
two-dimensional channels of constant width except channels of
constant curvature, which have no bound state. The quantum
effective potential that appears as a result of the presence of
curvature in low-dimensional systems has been studied also
in\cite{Enc*98, Enc*03, Kaplan, Schult, Carini} and
in\cite{Dan*05} where it was shown that a charged quantum particle
trapped in a potential of quantum nature due to bending of an
elastically deformable thin tube travels without dissipation like
a soliton. Surprisingly, the twist of a strip plays a role of a
magnetic field and is responsible for the appearance of localized
states and an effective transverse electric field thus reminisce
the quantum Hall effect\cite{Dan*04}.

It is possible to produce very narrow two-dimensional conducting
surfaces which allow electrons to propagate in the channel formed
by their boundaries, but require the electron wave function to
vanish on these boundaries. Furthermore, the effects induced by a
curved surface on the distribution of quantum particles are not
fully understood. In this paper, we study simple rotationally
invariant surfaces of varying curvature to gain a broader
understanding of the interaction between quantum particles and
curvature and possible physical effects.

The results of this paper are based on the exploration of the
properties of the Schr\"odinger equation on a sub-manifold of
$\mathbb{R}^3$\cite{Jensen,daCost*81}. Following da Costa an
effective potential appears in the Schr\"odinger  equation which
has the following form:
\begin{eqnarray}\label{eq:3}
V_s(q_1,q_2)=-\frac{\hbar^2}{2m}(M^2-K)=-\frac{\hbar^2}{8m}(k_1-k_2)^2,
\end{eqnarray}
where $m$ is particle's mass, $\hbar$--Plank's constant; $q_1$ and
$q_2$ are the generalized coordinates on the surface; $k_1$ and
$k_2$ are the principal curvatures of the surface and
$M=(k_1+k_2)/2$ and $K=k_1 k_2$ are the Mean and the Gauss
curvatures respectively.

The presence of the Mean curvature (which cannot be obtained from
the metric tensor and its derivatives alone) in (\ref{eq:3})
results in an important consequence: $V_s(q_1,q_2)$ is not the
same for two isometric surfaces. da Costa notes\cite{daCost*81}:
''thus independent of how small the range of values assumed for
$q_3$ (the third coordinate which measures the distance to the
surface along the normal vector), the wave function always moves
in three-dimensional portion of space, so that the particle is
"aware" of the external properties of the limit surface''. The
particle is also ''aware'' of the manner in which it is confined
to move to that limit surface\cite{Kaplan,Mitchell}. In view of
this, solving the inverse problem, we construct a deformed surface
which corresponds to an effective free one dimensional motion. We
also report the existence of a circular strip surface creating
conditions for a zero angular momentum particle to bind in a
harmonic potential.

Let us take a rotationally invariant surface $\vec{r}(q_1,q_2)$,
parameterized in Cartesian coordinates in Monge fashion:
\begin{equation}\label{eq:6}
\vec{r}(q_1,q_2)=\vec{r}(\rho, \phi)=(\rho \cos{\phi}, \rho
\sin{\phi}, f(\rho)),
\end{equation}
where $\rho \in [0,\infty)$ and $\phi \in [0, 2 \pi]$. The area
element is
\begin{equation}
ds^2=\rho^2 d \phi^2 + \left(1+\dot{f}(\rho)^2 \right)d \rho^2
\end{equation}
and the determinant of the metric is given by
\begin{equation}\label{eq:g}
g=\rho^2 \left(1+\dot{f}(\rho)^2 \right),
\end{equation}
where hereafter the dot represents derivative with respect to
$\rho$. From the first and the second fundamental forms of this
surface the following expressions for the principal curvatures are
obtained
\begin{eqnarray}\label{eq:8}
k_1(\rho)&=&\ddot{f}(\rho) \left(1+\dot{f}(\rho)^2
\right)^{-3/2},\\\label{eq:8a}
 k_2(\rho)&=&
\dot{f}(\rho) \left[\rho^2 \left(1+\dot{f}(\rho)^2
\right)\right]^{-1/2}.
\end{eqnarray}

If the surface is not a plane but it is asymptotically planar and
cylindrically symmetric then the Schr\"odinger operator can have
at least one isolated eigenvalue of finite multiplicity which
guarantees the existence of a geometrically induced bound
state\cite{Ex*01}. Looking for stationary modes in polar
coordinates in which the rotational invariance of the surface
(\ref{eq:6}) is obvious, we separate the variables $\chi_{\rm
t}(\rho, \phi, t)=\exp{(- {\rm i} E t/\hbar)} \exp{({\rm i}
\boldsymbol{\rm m} \phi)} \psi(\rho)$ (here $\chi_{\rm t}$ is the
tangential to the surface part of the wave
function\cite{daCost*81}) to end up with a quasi-one-dimensional
Sturm-Liouville  equation for the $\rho$--dependent part of the
wave function:
\begin{widetext}
\begin{eqnarray}\label{eq:13}
  \frac{1}{\rho \sqrt{1+\dot{f}^2}}
 \partial_{\rho}\left(\frac{\rho}{\sqrt{1+\dot{f}^2}}
 \partial_{\rho}
 \right)
 \psi_{\boldsymbol{\rm
m}}(\rho)- \frac{\boldsymbol{\rm m}^2}{\rho^2}
\psi_{\boldsymbol{\rm m}}(\rho) + \frac{2mE}{\hbar^2}
\psi_{\boldsymbol{\rm m}}(\rho)
 = \frac{2m}{\hbar^2} V_s(\rho)\psi_{\boldsymbol{\rm
m}}(\rho),
\end{eqnarray}
\end{widetext}
where $V_s(\rho)$ is given by (\ref{eq:3}) with (\ref{eq:8}) and
(\ref{eq:8a}). The normalization condition in $\rho$--space is
\begin{equation}\label{eq:P_rho}
2\pi \int_0^{\infty} \left|\psi_{\boldsymbol{\rm m}}(\rho)
\right|^2 \sqrt{g} d \rho=1
\end{equation}
with $g$ given by (\ref{eq:g}).

Due to the cylindrical symmetry and the conservation of the
$z$--component of the angular momentum, we may reduce the problem
to a one-dimensional equation for each angular momentum quantum
number $\boldsymbol{\rm m}$ along the Euclidean line length along
the geodesic on the surface with fixed $\phi$. Introducing the
changes\cite{CourHilb}
\begin{eqnarray}\label{eq:x}
  && x=\int_0^\rho \sqrt{1+\dot{f}^2(\rho')} d \rho',\\
&& \psi_{\boldsymbol{\rm m}}(\rho)= F_{\boldsymbol{\rm
m}}(x)/\sqrt{\rho},
\end{eqnarray}
we obtain for the function $F_{\boldsymbol{\rm m}}(x)$ a
one-dimensional Schr\"odinger equation which is the Liouville
normal form of (\ref{eq:13}):
\begin{equation}\label{eq:F}
-\frac{d^2}{dx^2} F_{\boldsymbol{\rm
m}}(x)+\left[W_{\boldsymbol{\rm m}}(x)-k^2 \right]
F_{\boldsymbol{\rm m}}(x)=0.
\end{equation}
Here we have introduced the wave vector instead of the energy
$\kappa^2=2mE/\hbar^2$. The geometrical properties of the surface
determine the quantum effects through the geometry dependent term
$W_{\boldsymbol{\rm m}}(x)$ in equation (\ref{eq:F}):
\begin{eqnarray}\label{eq:W}
W_{\boldsymbol{\rm m}}\left[x(\rho) \right] = -\frac14 k_1^2(\rho)
+ \frac{\boldsymbol{\rm m}^2-1/4}{\rho^2},
\end{eqnarray}
where $k_1$ is given by (\ref{eq:8}). The normalization condition
in $x$--space is
\begin{equation}\label{eq:P_x}
2\pi \int_0^{\infty} \left|F_{\boldsymbol{\rm m}}(x)   \right|^2 d
x=1.
\end{equation}

The term in $W_{\boldsymbol{\rm m}}$, proportional to
$\boldsymbol{\rm m}^2$, is the potential that describes the
familiar centrifugal force. Less familiar is the negative
correction term $-1/4$ which results not from the angular motion
but from the radial motion (and can be traced back to the radial
derivatives in the Laplacian expressed in the associated with the
surface coordinates (\ref{eq:6})). This is a coordinate force that
comes from the reduction of space from three to two dimensions and
was called {\it quantum anti-centrifugal force} by Cirone \etal
\cite{cirone*01} because it possesses binding power due to quantum
mechanics. Similar situation for the free radial motion of a
particle was noticed in\cite{berry*73}. This contribution is
strengthened by the binding curvature induced potential $-k_1^2
/4$, a geometric force.

An interesting situation (that may have practical application)
unfolds when we consider a Gaussian bump which appears often when
considering deformations of a surface with the following profile:
\begin{equation}\label{eq:12a}
f(\rho)=-A_0\exp\left(-{\rho^2}/{\sigma_0^2}\right),
\end{equation}
where $\sigma_0$ is the dispersion of the profile and $A_0 > 0$ is
its depth. Thus prior to the attempt to search for a solution of
(\ref{eq:F}) with (\ref{eq:W}) we would like to discuss the
physics.

The potential $W_{\boldsymbol{\rm m}}$ is repulsive for particles
with non-vanishing angular momentum ($\boldsymbol{\rm m} \neq 0$)
and no solutions with negative energy exist.

The effect from different contributions in the potential
$W_{\boldsymbol{\rm m}}$  stands out most clearly for particles
with zero angular momentum, that is, $\boldsymbol{\rm m}=0$. These
are attracted to the origin and are found in a ring-shaped region
around the axis of symmetry, while all particles with
$\boldsymbol{\rm m} \neq 0$ are repelled from the center. It is
clear that the bump not only introduces scale but also breaks the
symmetry of the $\mathbb{R}^2$ plane introducing a natural origin
-- its center.

\begin{figure}[ht]
\includegraphics[scale=0.20]{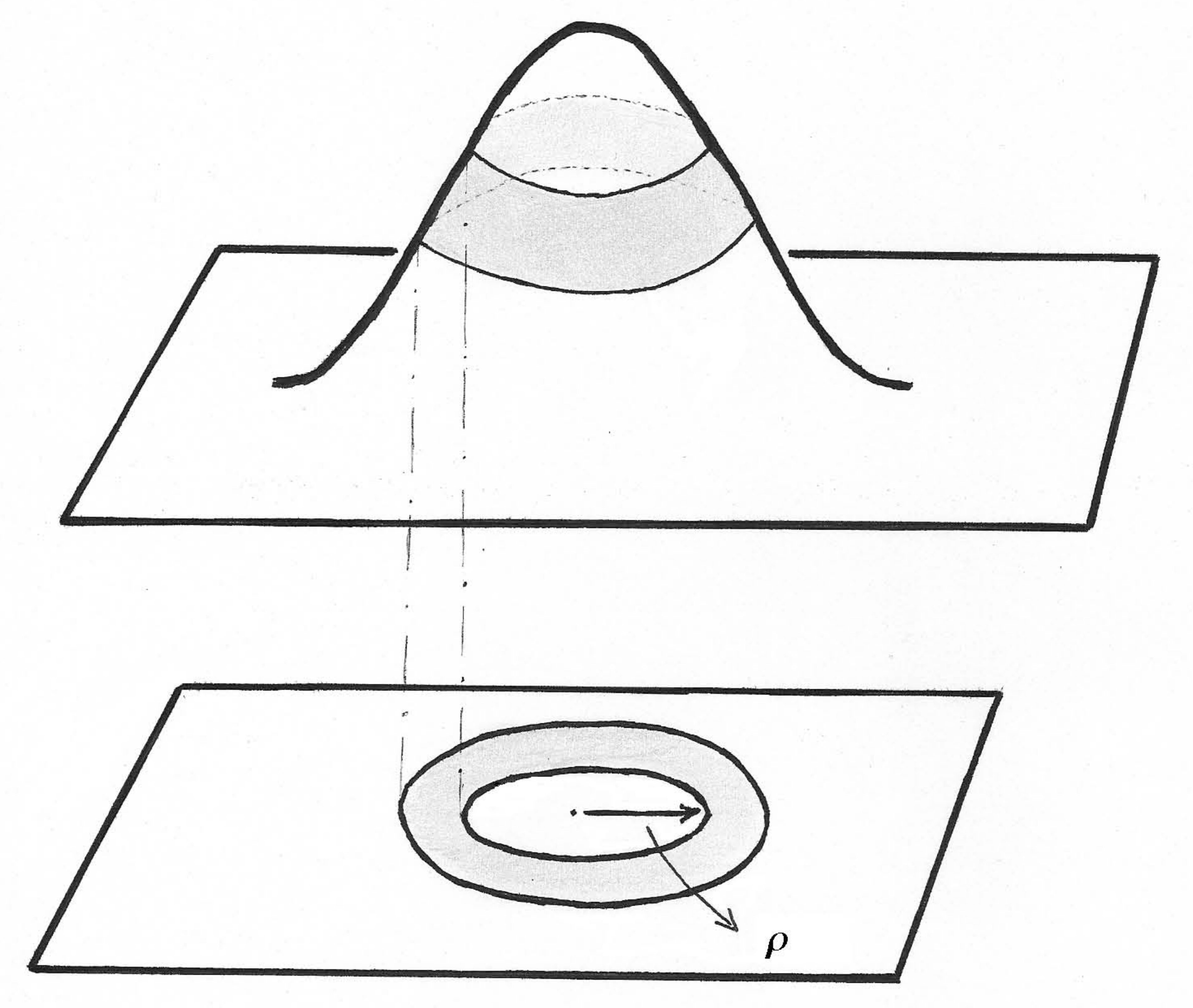}
\caption{\label{fig:1} A stretched strip on a bump and its
downward projection on a flat plane.}
\end{figure}

The Gaussian bump (\ref{eq:12a}) also introduces a characteristic
length on the surface. By vertically stretching the surface, the
amount of area in the stretched strip is increased with respect to
its projection on the plane which is depicted in Figure
(\ref{fig:1}). As a result, the quantum particle would rather
remain localized when it is situated near the maximal stretch of
the surface thus introducing a scale of the order of the
dispersion $\sigma_0$ of the profile (\ref{eq:12a}). This can
easily be seen when looking at the Heisenberg uncertainty relation
$\delta p \delta x \sim \hbar$: $\delta x$ is bigger in the
stretched area. Hence, it is energetically favorable the quantum
particle migrates to the stretched area of the bump thus creating
charge distribution if it itself is charged. This charge density
is exactly depicted by the probability density of finding the
particle in a specific place on the surface.

Since we are interested in bound states now we consider only
particles with $\boldsymbol{\rm m}=0$, since only for them the
potential $W_{\boldsymbol{\rm 0}}$ is negative and they will bind
with different quantized energies. Equating the kinetic and
potential energy terms we provide a rough estimate of the order of
magnitude of the bound state energies, namely
$E<-\hbar^2/2m\sigma_0^2$. Later on in the text we will see that
this estimate is correct up to a geometry dependent factor
$A_0^2/\sigma_0^2$, which measures the flatness of the profile.

Obtaining a solution to equation (\ref{eq:F}) is not an easier
task than solving the original one. Nevertheless, we can be sure
that bound states do exist since we have transformed the problem
to a one-dimensional one with a negative potential\cite{LL}. Using
an approximation we can transform this equation into simpler one
thus providing us with a chance to generate an approximate wave
function. The behaviour of the wave function in the vicinity of
the origin, that is $\rho \to 0$ can be considered as a starting
point in our consideration. In that particular limit $x \approx
\rho$ due to (\ref{eq:x}) and $\lim_{\rho \to 0}\dot{f}=0$ (for
the Gaussian bump) for the potential $W_{\boldsymbol{\rm 0}}(x)$
we have
\begin{equation}\label{eq:W_0}
W_{\boldsymbol{\rm 0}}\left[x \right]= -\frac{1}{4
x^2}-\frac{A_0^2}{\sigma_0^2}\frac{1}{\sigma_0^2}.
\end{equation}
Equation (\ref{eq:F}) with the above potential possesses a stable
negative energy solution with
$k'^2={2m|E|}/{\hbar^2}-{A_0^2}/{\sigma_0^4}$ provided
${2m|E|}/{\hbar^2}>{A_0^2}/{\sigma_0^4}$. Here
${A_0^2}/{\sigma_0^2}$ is a measure of the flatness of the
Gaussian bump. Due to the sign change of the energy the ordinary
Bessel functions $J_{\boldsymbol{\rm 0}}$ and $Y_{\boldsymbol{\rm
0}}$ which solve equation (\ref{eq:F}) with the above potential
(\ref{eq:W_0}) turn into the modified Bessel functions
$I_{\boldsymbol{\rm 0}}$ and $K_{\boldsymbol{\rm
0}}$\cite{Abr&Steg}. Since the modified Bessel function
$I_{\boldsymbol{\rm 0}}$ increases exponentially for large
distances, whereas $K_{\boldsymbol{\rm 0}}$ decreases, the
boundary conditions imposed by the asymptotic flatness of the
Gaussian bump and the need for a squared integrable wave function,
we select $K_{\boldsymbol{\rm 0}}$. Thus the solution in the
vicinity of $x \to 0$ is expressed with the modified Bessel
function
$$F_{{\boldsymbol{\rm 0}}}(x) \approx \sqrt{x} K_{\boldsymbol{\rm
0}}(k' x).$$

Having deduced the correct behaviour of the wave function in the
vicinity of the origin we can try to solve equation (\ref{eq:F})
with the ansatz
\begin{equation}
F_{\boldsymbol{\rm 0}}(x)=\exp\left[\frac{i}{\hbar}S(x)\right]
\sqrt{x} K_{\boldsymbol{\rm 0}}(k' x).
\end{equation}
Using the fact that $K_{\boldsymbol{\rm 0}}$ satisfies the Bessel
equation we end up with an approximate equation for $S(x)$
\begin{equation}\label{eq:S}
\frac{i}{\hbar} S^{''}  = \frac{1}{\hbar^2} \left(S^{'} \right)^2
+\frac{i}{\hbar} \frac{S^{'}}{x}-\left[
  \frac{1}{4} \frac{\ddot{f}^2}{\left(1+\dot{f}^2\right)^{3}}
  -\frac{A_0^2}{\sigma_0^4}\right] ,
\end{equation}
where hereafter $(\; ^{'})$ denotes differentiation with respect
to $x$. In the derivation of (\ref{eq:S}) we kept in mind that $x
\approx \rho$ and approximated $1/\rho^2$ with $1/x^2$ in the
potential (\ref{eq:W}). We have also approximated $ {d
K_{\boldsymbol{\rm 0}}}/{dx} \approx -K_{\boldsymbol{\rm 0}}/ x $
which preserved the behaviour near the origin.

Next we expand $S(x)$ in series with increasing powers of the
small parameter $\hbar$ {\it \`a la} WKB type expansion
\begin{equation*}
S(x)=S_0(x)+i \hbar S_1(x)+ \ldots
\end{equation*}
and obtain for the first two functions in the expansion the
equations
\begin{eqnarray}
S_0^{'} &=& 0 \quad \Rightarrow \quad S_0={\rm const},\\
\nonumber S_0^{''} &=& 2 S_0^{'} S_1^{'} + S_0^{'}/x \quad
\Rightarrow \quad \forall S_1^{'},\\\label{eq:S_1} S_1^{''} &=&
(S_1^{'})^2 + S_1^{'}/x - \left(
  \frac{ A_0^2}{\sigma_0^4}
  -\frac{1}{4}
  \frac{\ddot{f}^2}{\left(1+\dot{f}^2\right)^{3}}\right).
\end{eqnarray}
Solving (\ref{eq:S_1}) we assume that i.) $(S_1^{'})^2 \gg
S_1^{''}$ and ii.) $(S_1^{'})^2 \gg S_1^{'}/x$ to obtain
\begin{equation}\label{eq:sol_S_1}
 S_1^{'} = \pm \sqrt{
  \frac{ A_0^2}{\sigma_0^4}
  -\frac{1}{4}
  \frac{\ddot{f}^2}{\left(1+\dot{f}^2\right)^{3}} }.
\end{equation}
From the form of the solution we see that the assumptions i.) and
ii.) we made in deriving it are justified away from the origin.
Since we already dealt with the behaviour of the wave function in
the vicinity of that point we write
\begin{equation}\label{eq:F_fin}
F_{\boldsymbol{\rm 0}}(x)=\frac{e^{i S_0/\hbar}}{\sqrt{2 \pi N^2}}
\sqrt{x} K_{\boldsymbol{\rm 0}}(k' x) \exp\left[-\int_0^x S_1^{'}
dx \right],
\end{equation}
where we have chosen the $+$ sign solution of (\ref{eq:sol_S_1}),
ameliorating the vanishing behaviour of the wave function at
infinity. The norm $N^2$ is calculated using (\ref{eq:P_x}). The
logarithmic divergence at the origin of the modified Bessel
function of second kind $K_{\boldsymbol{\rm 0}}$ is regularized by
the extra factor proportional to $\sqrt{x}$ in terms of the
Euclidean line length on the surface, making the probability
density to find the particle at the origin vanishing. Moreover,
since (\ref{eq:F_fin}) vanishes exponentially for large distances,
the radial probability displays a maximum close to the origin,
which is exactly the behaviour of the expectation value that we
already deduced from the considerations using Heisenberg's
principle.

As a result, particles with $\boldsymbol{\rm m}=0$ concentrate in
a ring shaped region on the surface of the Gaussian bump around
the axis of symmetry.

\begin{figure}[ht]
\includegraphics[scale=0.38]{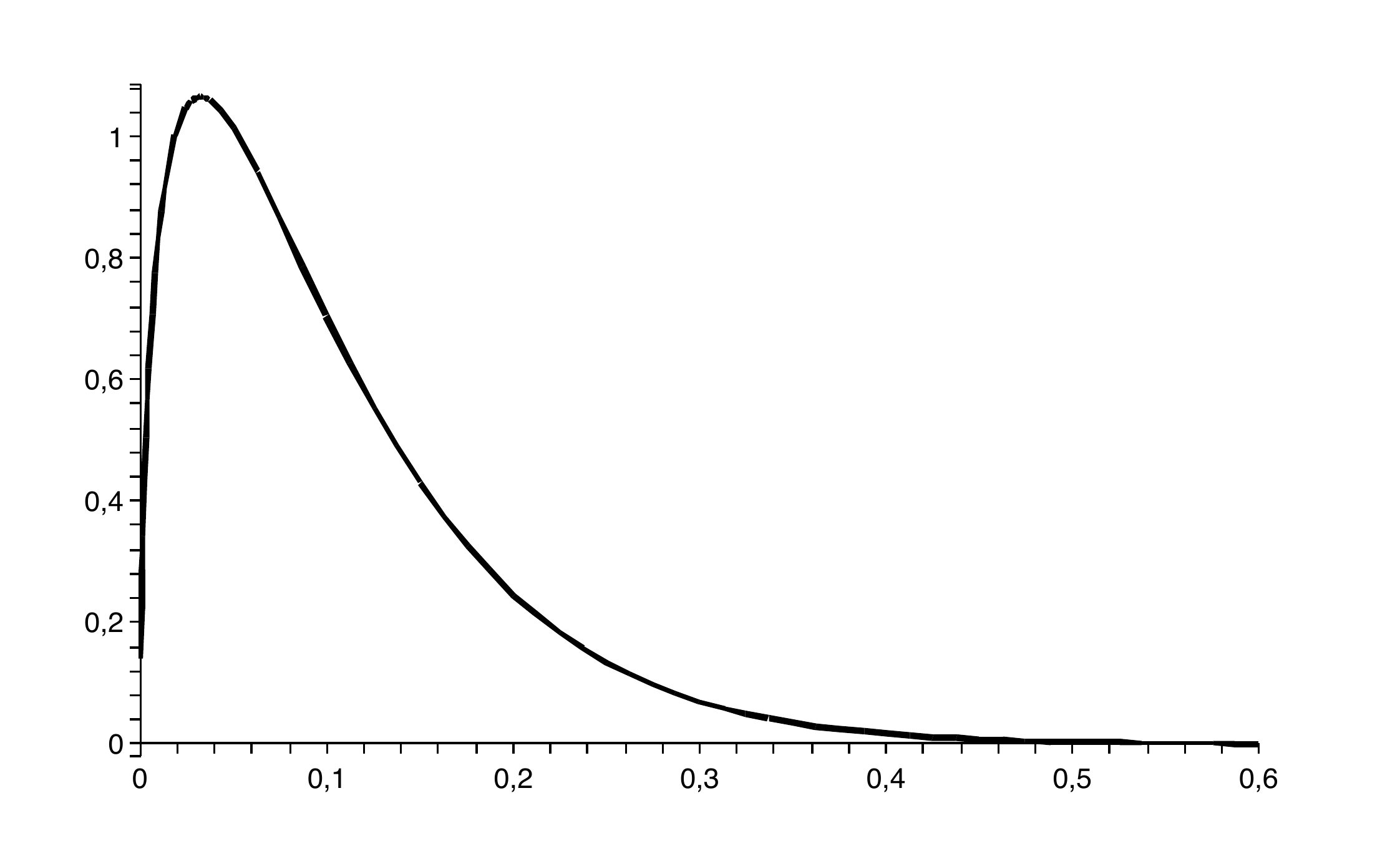}
\caption{\label{fig:2} The probability density
$|\psi_{\boldsymbol{\rm 0}}(\rho)|^2\sqrt{g}$ (vertical axis) to
find a particle between $\rho$ and $\rho + d \rho$ (horizontal
axis) for the Gaussian bump. Here we have set $A_0=\sigma_0=1$ and
$k'=5$. The initial phase $S_0=0$. }
\end{figure}

Figure (\ref{fig:2}) depicts thus generated probability density
$|\psi_{\boldsymbol{\rm 0}}(\rho)|^2\sqrt{g}$ of finding the
particle in a state with angular momentum $\boldsymbol{\rm m}=0$
between $\rho$ and $\rho + d \rho$ in terms of the distance from
the axis of symmetry and clearly shows that the probability of
finding the quantum particle is biggest at the stretched area of
the bump as already discussed. This shows that the approximations
which lead to (\ref{eq:F_fin}) are fully justified. They also show
the way to a proper test function for a run of the variational
method to determine the ground state energy and wave function,
which must not have nodes -- a property clearly visible in
(\ref{eq:F_fin}). This is a matter of a future study.

Here we do not present an estimate of the structure of the energy
spectrum except the already stated fact -- the discrete spectrum
starts at $E < - \hbar^2A_0^2/2m\sigma_0^4$, value which is
determined by the geometrical extensions of the Gaussian bump.
Over that value the motion of a particle with $\boldsymbol{\rm
m}=0$ is classically infinite and hence the wave function is
supposed to be in the form of a plane wave. What is, at least, the
approximate expression of that plane wave?

We propose to look for solutions of (\ref{eq:F}) in the form of
\begin{equation}
F_{\boldsymbol{\rm 0}}^{\rm cont}=f_{\boldsymbol{\rm
0}}(x)\exp{\left( \pm i \int Q_{\boldsymbol{\rm 0}} dx \right)},
\end{equation}
where $Q_{\boldsymbol{\rm 0}} \approx \sqrt{ k_1^2/4 + 1/4x^2 +
k^2 }$. Here $f_{\boldsymbol{\rm 0}}(x)$ is fast changing
function, such that $F_{\boldsymbol{\rm 0}}^{''}(x) \approx
\left(-Q_{\boldsymbol{\rm 0}}^2  \pm 2i Q_{\boldsymbol{\rm 0}}
f_{\boldsymbol{\rm 0}}^{'}/f_{\boldsymbol{\rm 0}} \pm i
Q_{\boldsymbol{\rm 0}}^{'} \right)F_{\boldsymbol{\rm 0}}$. From
here and (\ref{eq:F}) we see that
\begin{equation}
 \frac{ f_{\boldsymbol{\rm
0}}^{'}}{f_{\boldsymbol{\rm 0}}} \approx -\frac12 \frac{
Q_{\boldsymbol{\rm 0}}^{'}}{Q_{\boldsymbol{\rm 0}}} \quad
\Rightarrow \quad f_{\boldsymbol{\rm 0}} \sim
1/\sqrt{Q_{\boldsymbol{\rm 0}}}.
\end{equation}
Hence we write
\begin{equation}
F_{\boldsymbol{\rm 0}}^{\rm cont}(x)=\frac{\exp{\left( \pm i \int
\sqrt{k_1^2/4 + 1/4x^2 + k^2 } dx \right)}}{\left(k_1^2/4 + 1/4x^2
+ k^2 \right)^{1/4}},
\end{equation}
where the plane wave exhibits a node in the origin.

The energy spectrum for particles with non-vanishing angular
momentum is continuous and starts at $E=0$. Good approximate
solutions of (\ref{eq:F}) are the ordinary Bessel functions
$\sqrt{x} J_{\boldsymbol{\rm m}}(k x)$ and $\sqrt{x}
Y_{\boldsymbol{\rm m}}(k x)$, whose argument is the Euclidean line
length on the surface.

\medskip

Let us now turn our attention to the inverse problem or
equivalently the question "What rotationally invariant surface
leads to an effective geometry induced potential
$W_{\boldsymbol{\rm m}}$ that equals prescribed negative function
$-U[x(\rho)]$, where $U \geq 0$ for $\forall \rho$?" (negative
because we are primary interested in bound states). This question
is of particular interest since for certain classes of negative
potentials $-U$ we already know the exact wave functions which can
readily be used in revealing the particle's distribution on the
surface. The solution of the inverse problem goes through the
recognition of its equivalence with the following differential
equation (see equations (\ref{eq:W}) and (\ref{eq:8})):
\begin{eqnarray}\label{eq:29}
\frac14 \frac{\ddot{f}(\rho)^2}{ \left(1+\dot{f}(\rho)^2
\right)^{3}}=U(\rho)+ \frac{\boldsymbol{\rm m}^2-1/4}{\rho^2}.
\end{eqnarray}

From equations (\ref{eq:8}) and (\ref{eq:29}) one can easily
deduce a condition on $k_1(\rho)$ in order to have a binding
potential for a particle with $\boldsymbol{\rm m} \neq 0$, i.e. $$
k_1^2 > \frac{4 \boldsymbol{\rm m}^2-1}{\rho^2}.$$ For smooth
surfaces at $\rho=0$ this means that there will be only strips
where this condition holds. For flat surfaces where $k_1=0$ this
condition implies that only particles with $\boldsymbol{\rm m}=0$
bind, a fact previously noticed in\cite{cirone*01, berry*73}. Now we try to
solve equation (\ref{eq:29}) using substitution
$\dot{f}=\sinh{(\omega)}$ (here $\omega=\omega(\rho)$) to end up
with the linear equation
\begin{eqnarray}\label{eq:30}
\frac12 \frac{d \tanh(\omega)}{d \rho} =\pm \sqrt{ U(\rho)+
\frac{\boldsymbol{\rm m}^2-1/4}{\rho^2} },
\end{eqnarray}
which can be solved thus yielding a result for $\dot{f}(\rho)$. We
integrate to obtain the profile of the surface
\begin{eqnarray}\label{eq:f}
f(\rho)= \pm \int^{\rho}_{\rho_1}
\frac{|\mathcal{A}|}{\sqrt{1-\mathcal{A}^2
 } } d \rho',
\end{eqnarray}
where
\begin{equation}\label{eq:A}
\mathcal{A}(\rho) =\pm 2 \int^{\rho}_{\rho_0} \sqrt{ U(\rho') +
\frac{\boldsymbol{\rm m}^2-1/4}{{\rho'}^2} } d \rho'.
\end{equation}
Here $\rho_0$ and $\rho_1$ are constants of integration and are to
be determined by the boundary conditions due to the behavior of
the function $U$ representing the potential we want to model.
Since ${f}(\rho)$ takes only real values (the same is true for all
of its derivatives, i.e. ${d^{\rm n} f}/{d \rho^{\rm n}} \in
\mathbb{R}$ for ${\rm n}=0,1,\ldots$) as a function describing the
profile of a surface we impose $0 < |\mathcal{A}| < 1$. The
$\mathcal{A}=0$ case is realized by a flat surface. Using the
theorem of the mean value in (\ref{eq:A}) and the above inequality
we obtain
\begin{equation}
\rho_0 < \rho  < \rho_0 + \frac12 \left| U(\xi) +
\frac{\boldsymbol{\rm m}^2-1/4}{{\xi}^2} \right|^{-1/2},
\end{equation}
where the point $\xi \in [\rho_0, \rho]$. In that manner we show
that for generic angular momentum $\boldsymbol{\rm m}$ and
non-zero potential $U$ the corresponding rotational surface
creating this potential and allowing a bound state of a quantum
particle with ${\boldsymbol{\rm m}} \neq 0$ exists only in a
ribbon, a circular strip around the axis of symmetry.

Let us also note that for $\boldsymbol{\rm m} \neq 0$ the bump (a
macroscopic structure) has a magnetic moment, i.e. the macroscopic
deformation of the surface acquires quantum number. Indeed the
probability density current $\vec{J}$ (${\rm div} \vec{J}=0$)
associated with the wave function $\chi_{\rm t}$ is given by
\begin{equation*}
\vec{J}=\left(
  J_{\phi}, J_{\rho}, J_z \right)=\frac{\hbar}{m}\left(
    \boldsymbol{\rm m} \frac{ |\psi_{\boldsymbol{\rm m}}|^2}{\rho},  {\rm Re} \frac{  \psi_{\boldsymbol{\rm m}}^{\ast}
   \partial_{\rho}\psi_{\boldsymbol{\rm m}}
   }{i\sqrt{1+\dot{f}^2}  }, 0
   \right),
\end{equation*}
where $J_{\phi}, J_{\rho}$ and $J_{z}$ are the components of the
vector $\vec{J}$ in the orthonormal right-handed triad
$(\vec{e}_{\rho}, \vec{e}_{\phi},\vec{e}_{z})$, has nonzero and
quantized with $\boldsymbol{\rm m}$ circulation along the
circumference of the bump.

Now let us give a couple of examples:

\begin{figure}[t]
\includegraphics[scale=0.45]{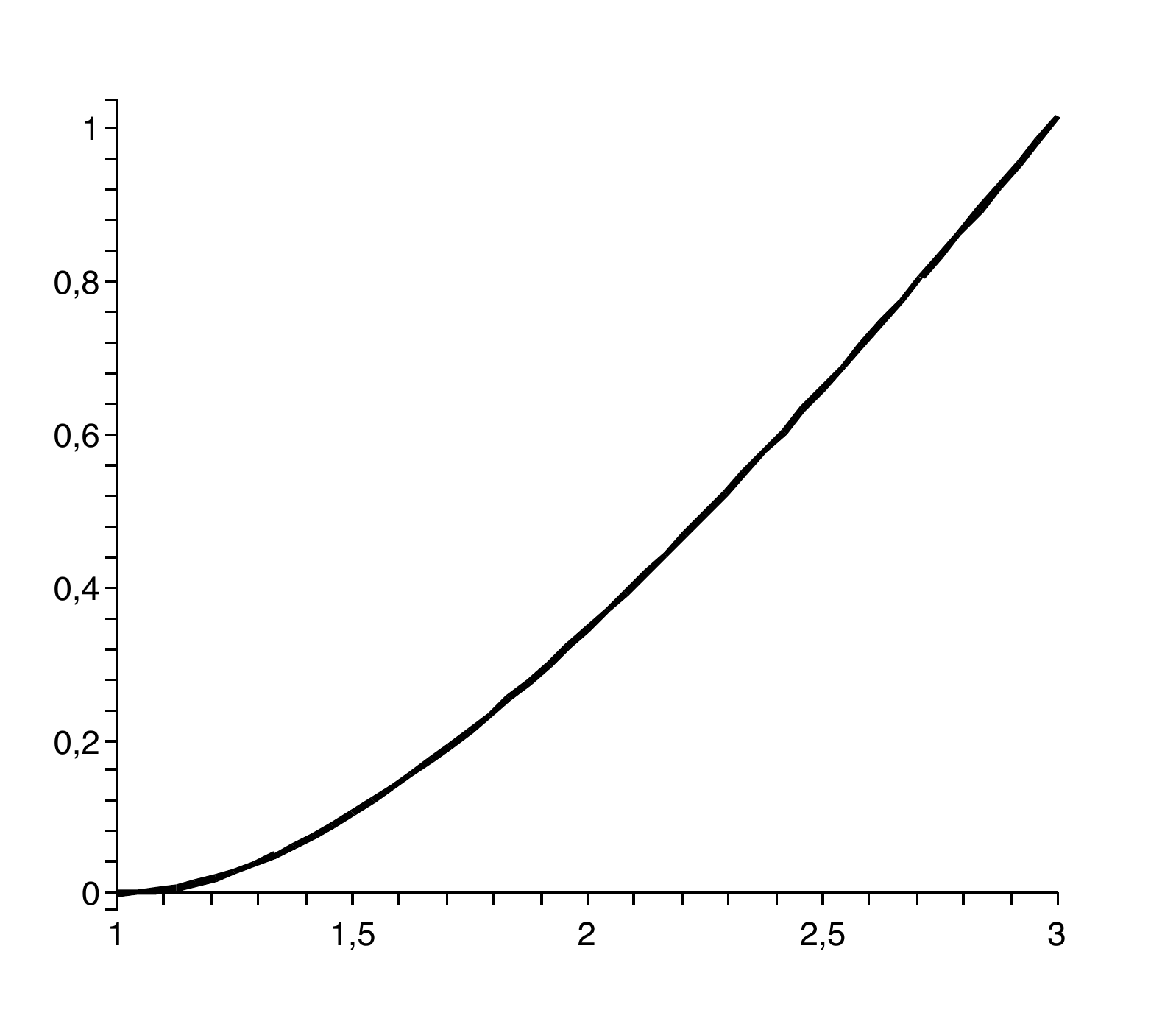}
\caption{\label{fig:3} The surface $f(\rho)/\rho_0$ (vertical
axis) with a cusp on which a particle with $\boldsymbol{\rm m}=0$
moves freely. On the horizontal axis is plotted $\rho/\rho_0$. At
infinity that surface tends to a cone.}
\end{figure}

\begin{itemize}

\item $U=0$, i.e. free motion; From (\ref{eq:A}) and (\ref{eq:f})
it follows that
\begin{equation}\label{eq:f_free}
f^{\rm free}_{\boldsymbol{\rm m}}(\rho)= \pm  \rho_0
\int^{\rho/\rho_0}_{\rho_1/\rho_0} \frac{ \sqrt{|4 \boldsymbol{\rm
m}^2-1|} \left|\ln\left(\rho'
\right)\right|}{\sqrt{1-(4\boldsymbol{\rm m}^2-1) \; \ln^2
\left(\rho'\right)
 } } d \rho'.
\end{equation}
Here $\rho_0$ determines the characteristic scale of the surface
on which we consider this free quantum problem. Its magnitude can
be chosen in $[\mu m]$ or $[n m]$ depending on the scale that we
want to model. If we consider the shape of a surface allowing a
free motion of a particle with zero angular momentum, that is
$\boldsymbol{\rm m}=0$, we can integrate (\ref{eq:f_free}) with
$\rho_1=\rho_0$ to produce Figure (\ref{fig:3}). That surface is
asymptotically flat as it tends to a cone $f^{{\rm
free}}_{\boldsymbol{\rm 0}}(\rho) \to \rho$ as $\rho \to \infty$.

Next we consider the strip of a surface allowing a free motion of
a particle with non-zero angular momentum, that is
$\boldsymbol{\rm m} \neq 0$. We can impose the condition
$1>(4\boldsymbol{\rm m}^2-1) \; \ln^2 \left(\rho/\rho_0\right)$ to
find the extensions of the strip
\begin{equation}
\rho_0 < \rho < \rho_0 \exp{\left[(4\boldsymbol{\rm m}^2-1)^{-1/2}
\right]}.
\end{equation}
Since the problem is defined on a strip, we can quantize it in the
usual finite volume method, which would lead to standing wave
solutions on the surface. Their energy is
\begin{equation}
E_{n}^{\boldsymbol{\rm m}} (\rho_0)= \frac{2 \pi^2 \hbar^2 n^2}{m
\rho_0^2 \left[e^{(4\boldsymbol{\rm m}^2-1)^{-1/2} } -1 \right]^2
}.
\end{equation}

\item $U=\omega^2 \rho^2$, i.e. harmonic oscillator potential;
From (\ref{eq:A}) it follows that
\begin{equation}\label{eq:A_har}
\mathcal{A}^{{\rm harm}}_{\boldsymbol{\rm m}}(\rho) =\pm 2
\int^{\rho}_{\rho_0} \sqrt{ \omega^2 {\rho'}^2+
\frac{\boldsymbol{\rm m}^2-1/4}{{\rho'}^2} } d \rho'.
\end{equation}
Let us consider two cases consequently:

I.) The ${\boldsymbol{\rm m}}=0$ case yields for (\ref{eq:A_har})
the result
\begin{eqnarray}
\nonumber \mathcal{A}^{{\rm harm}}_{\boldsymbol{\rm 0}}(\rho) &=&
\pm \frac12 \left\{ \sqrt{4 \omega^2 \rho^4-1 } \right.\\
&&+\left. \arctan{\left[ \left(4 \omega^2 \rho^4-1 \right)^{-1/2}
\right]}-\frac{\pi}{2} \right\},
\end{eqnarray}
where we have set the value of $\rho_0=1/\sqrt{2|\omega| }$. From
the requirement $0 < |\mathcal{A}^{{\rm harm}}_{\boldsymbol{\rm
0}}| < 1$ we obtain the following estimate
\begin{equation}
\rho < \frac{\left( 1 +  (2+ \varepsilon )^2
\right)^{1/4}}{\sqrt{2|\omega|} }  \approx \frac{5^{1/4} +
{\varepsilon}/{5^{3/4}} }{\sqrt{2|\omega|} },
\end{equation}
where $\varepsilon={\pi}/{2} - \arctan{\left(4 \omega^2 \rho^4-1
\right)^{-1/2}} \ll 1$. Thus we find an expression for the
extensions of the circular strip of a rotational surface creating
harmonic potential and allowing a bound state of a particle with
vanishing angular momentum in a harmonic oscillator potential
\begin{equation}
\frac{1}{\sqrt{2|\omega|} } \leq \rho < \left( 5^{1/4}
  + \frac{\varepsilon}{5^{3/4}} \right)\frac{1}{\sqrt{2|\omega|} }.
\end{equation}

II.) The ${\boldsymbol{\rm m}} \neq 0$ case yields for
(\ref{eq:A_har}) the result
\begin{widetext}
\begin{eqnarray}\label{eq:A_har_m}
\pm \mathcal{A}^{{\rm harm}}_{\boldsymbol{\rm m}}(\rho) =
\sqrt{\omega^2 {\rho}^4+ {\boldsymbol{\rm m}^2-1/4} }
-\sqrt{\boldsymbol{\rm m}^2-1/4 }\; { \rm arctanh} {\left[
\left(1+\frac{\omega^2 \rho^4 }{\boldsymbol{\rm m}^2-1/4}
\right)^{-1/2} \right]}.
\end{eqnarray}
\end{widetext}
Next we impose the condition $0< |\mathcal{A}^{{\rm
harm}}_{\boldsymbol{\rm m}} | < 1$ and numerically find for
$\boldsymbol{\rm m}=1,2,3$ particles
\begin{eqnarray}
\nonumber && 1.071 < \sqrt{2 \omega} \rho < 1.602 \qquad
\boldsymbol{\rm
m}=1,\\
&& 1.602 < \sqrt{2 \omega} \rho < 1.957 \qquad \boldsymbol{\rm
m}=2,\\
\nonumber && 1.980 < \sqrt{2 \omega} \rho < 2.265 \qquad
\boldsymbol{\rm m}=3.
\end{eqnarray}

\end{itemize}

In conclusion we have calculated the quantum potential associated
with a Gaussian deformation of a two-dimensional plane. For that
simple surface deformation we have shown that a quantum binding
force of geometric origin attracts only particles with
$\boldsymbol{\rm m}=0$ to the origin. Further, we have solved the
inverse problem and have shown that surfaces in a form of a ribbon
allow bound states of particles with generic angular momenta. Thus
we speculate that a classical object (the ribbon) exhibits quantum
characteristics (the magnetic moment due to non-vanishing
quantized probability current circulation along the circumference)
acquired due to curvature. We have also found and depicted the
surface which corresponds to a free motion of a particle with
vanishing angular momentum.

\end{document}